\documentclass[conference,a4paper]{IEEEtran}

\usepackage{graphicx}
\usepackage{amsmath}
\usepackage{amssymb}
\usepackage{accents}
\usepackage{cite}

\newcommand\thickbar[1]{\accentset{\rule{.4em}{.8pt}}{#1}}

\newtheorem{definition}{Definition}
\newtheorem{remark}{Remark}

\IEEEoverridecommandlockouts
\begin{document}

\title{Energy Management in Wireless Communications with Energy Storage Imperfections}

\author{Sami Ak{\i}n\thanks{This work was supported by the German Research Foundation (DFG) -- FeelMaTyc (FI 1236/6-1)}\\
Institute of Communications Technology\\
Leibniz Universit\"{a}t Hannover\\
Email: sami.akin@ikt.uni-hannover.de}

\maketitle

%%% To be inactivated for final submission
%\thispagestyle{plain}
%\pagestyle{plain}
%%% To be inactivated for final submission

\begin{abstract}
In recent years, energy harvesting has taken a considerable attention in wireless communication research. Nonetheless, the stochastic nature of renewable energy sources has become one of the research problems, and energy storage has been proposed as a solution to deal with it. Initially, researchers regarded a perfect battery model without energy losses during storage because of its simplicity and compatibility in wireless communication analysis. However, a battery model that reflects practical concerns should include energy losses. In this paper, we consider an energy harvesting wireless communication model with a battery that has energy losses during charging and discharging. We consider energy underflows (i.e., the energy level falls below a certain threshold in a battery) as the energy management concern, and characterize the energy underflow probability and provide a simple exponential formulation by employing the large deviation principle and queueing theory. Specifically, we benefit from the similarity between the battery and data buffer models. We further coin the available space decay rate at a battery as a parameter to indicate the energy consumption performance. We further outline an approach to set the energy demand policy to meet the energy management requirements that rule the energy underflow probability as a constraint. We finally substantiate our analytical findings with numerical demonstrations, and compare the transmission performance levels of a transmission system with a battery that has energy losses and a transmission system that consumes the energy as soon as it is harvested.
\end{abstract}

\section{Introduction}\label{sec:introduction}
We can consider energy harvesting as a potentially feasible technique to provide tenable energy sources in wireless communications. Exploiting the energy in nature, we can set relatively less stringent power constraints for data transmission than we have in non-rechargeable battery-dependent wireless communication networks. We can also anticipate less human interventions in a network after its initialization, because there will be less number of battery changes. Nevertheless, there is the stochastic nature of renewable energy sources as one of the rudimentary problems to be handled before establishing a reliable communication system. Here, energy storage has been proposed as a solution to tackle the randomness in energy availability. However, one needs to consider energy storage efficiency, because there will be losses during the process of energy storing, e.g., energy losses while charging and discharging a battery, and energy leakage during storage. Therefore, we need an analytical framework that is simple, yet comprehensive, and can encapsulate the characteristics of a generic energy harvesting communication system with battery imperfections.

In order to formulate energy storage dynamics in wireless communications, researchers have established several mathematical models. While many of these models can reflect energy storing characteristics, queueing theory-based energy quantization models can easily fit into communication models and generally provide less-complicated analysis when invoked in wireless communications \cite{tadayon2013power}. Here, some researchers benefited from information-theoretic tools, and investigated performance levels under energy harvesting constraints, e.g., \cite{6736899,6381384,orhan2013throughput,ozel2012achieving}. Other researchers, considering a transmitter with a data buffer and a battery, studied the data-link layer performance, where their objective was to manage simultaneously the data traffic load and the stored energy, respectively, under delay and energy constraints \cite{srivastava2013basic,gong2014supporting,5992840,ozel2013optimal,sharma2016packet}. Furthermore, using the same mathematical model, some other researchers addressed the power allocation policies in batteries \cite{4211894,6202352}, and provided a stochastic modeling for these batteries \cite{3229100,Kansal:2004:PAT:1012888.1005714}.

Given a stochastic energy harvesting (arrival) process at a battery or any storage device, it is of importance to set an energy demand process that utilizes the stored energy as efficiently as possible. Moderate energy usage leads to energy overflows due to the limited battery size, and causes energy losses, while faster energy consumption results in energy outages, which induce transmission interruptions. Therefore, energy demand rates are essential metrics to take into account. Another constraint on using a battery is that we cannot drain all the energy from a battery but we should use a certain amount of the available energy in order to extend the battery life \cite{ghiassi2015toward}. In this case, rather than energy outages, controlling energy underflows (i.e., the energy level falls below a certain threshold in a battery) serves the purpose of energy management both to extend battery life and to prevent data transmission interruptions. In this context, the authors of \cite{zhang2015joint} regarded the energy underflow probability as the battery constraint, and attained the optimal constant power allocation policy that maximizes the effective capacity. Furthermore, we should also keep in mind energy losses during charging and discharging a battery. In other words, after harvesting the energy, we should carefully calculate how much energy we should store and how much energy should directly use for data transmission.

\begin{figure*}
	\centering
	\includegraphics[width=0.65\textwidth]{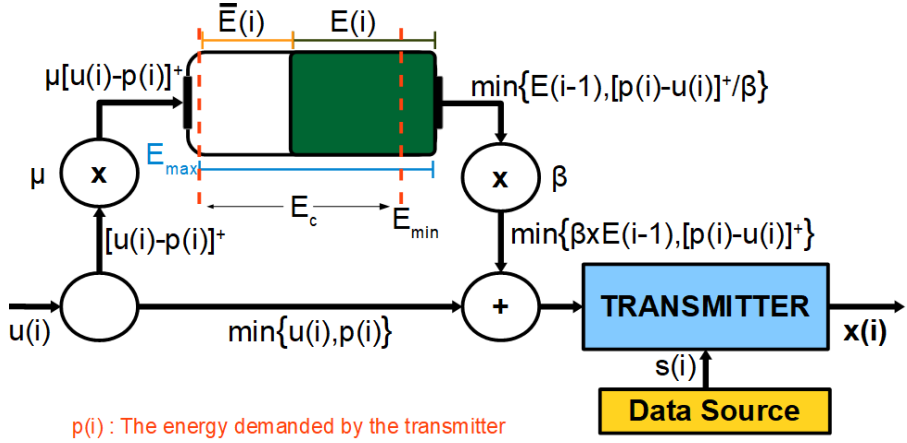}
	\caption{System model. $[\cdot]^{+}=\max\{\cdot,0\}$.}\label{res:system_model}
\end{figure*}
In this paper, we focus on energy management in an energy harvesting wireless communication system, where a transmitter harvests energy from the nature and stores it in a battery. We assume that the transmitter uses the entire harvested energy for data transmission. We also consider a limited-size battery model with energy losses during charging and discharging. We take advantage of the large deviation principle, and introduce an analytical framework that system designers can use in order to understand the performance levels in a general class of energy harvesting communication systems under the energy underflow probability constraint. We formulate a simple exponential approximation for the energy underflow probability, where we characterize the available space decay rate (i.e., the decay rate of the tail distribution of the available space to store more energy) in the battery as a measure of energy utilization. Furthermore, we assume that the transmitter initially sets an energy demand policy taking into account the energy underflow probability constraint and the energy arrival statistics, and then it sets the transmission rate based on the availability of the energy and the channel conditions. Finally, we compare the performance levels of an energy harvesting communication system with a battery that has energy losses, and a system without a battery, i.e., a system that uses the energy as soon as it is harvested. We also note that the difference between this study and \cite{akin2017energy} is in the battery models. Particularly, we consider a perfect battery model in \cite{akin2017energy}, i.e., there are no energy losses during energy storage, while we invoke an imperfect battery model with energy losses in this study. This is fundamental because the battery model with imperfections reflects more practical aspects.

\section{System Model}\label{sec:system_model}
\subsection{Energy Harvesting and Storage}
As seen in Figure \ref{res:system_model}, we consider a communication system composed of a battery, a transmitter and a data source. The system harvests energy from the nature, stores it in the battery, and then uses it for data transmission. $E_{\max}$ is the total storage capacity of the battery. $E_{\min}$ is the minimum necessary energy level in the battery. In other words, the system should avoid battery depletion as much as possible and not permit the energy level to fall below $E_{\min}$. Therefore, we can consider the difference between $E_{\max}$ and $E_{\min}$ as the feasible battery capacity, i.e., $E_{\text{c}}=E_{\max}-E_{\min}$. Moreover, assuming that the discretization precision is above an acceptable value, we model the energy arrival and demand processes as discrete-time processes. In particular, $E(i)$ and $\thickbar{E}(i)$ are the available energy and the available space to store more energy, respectively, at the end of the $i^{\text{th}}$ time frame. Hence, we have $E_{\max}=E(i)+\thickbar{E}(i)$. The transmitter demands $p(i)$ units of energy, and the system harvests $u(i)$ units of energy in the $i^{\text{th}}$ time frame. If the harvested energy, $u(i)$, is greater than the energy demanded, $p(i)$, the excess energy, i.e., $u(i)-p(i)$, is sent to the battery, and the battery is charged. On the other hand, if $u(i)$ is less than $p(i)$, the transmitter requests the rest of the energy, i.e., $p(i)-u(i)$, from the battery, and the battery is discharged. During charging and discharging, the battery loses energy. Specifically, $\mu[u(i)-p(i)]^{+}$ units of energy is stored in the battery during charging, where $0<\mu<1$ is the battery charging rate, and $[\cdot]^{+}=\max\{\cdot,0\}$. Furthermore, the battery sends $\frac{1}{\beta}[p(i)-u(i)]^{+}$ units of energy in order to meet the energy demand, i.e., $p(i)-u(i)$, when $u(i)<p(i)$, where $0<\beta<1$ is the battery discharging rate. If the battery does not have enough energy to cover the demand, i.e., $E(i-1)<\frac{1}{\beta}[p(i)-u(i)]^{+}$, it sends the entire available energy to the transmitter, where $E(i-1)$ is the amount of energy in the battery at the end of the $(i-1)^{\text{th}}$ time frame.

We can express the energy level in the battery at the end of the $i^{\text{th}}$ time frame as
\begin{align}\label{eq:actual_energy_level}
	E(i)=\min\bigg\{&\Big[E(i-1)+\mu[u(i)-p(i)]^{+}\nonumber\\
	&-\frac{1}{\beta}[p(i)-u(i)]^{+}\Big]^{+},E_{\max}\bigg\}.
\end{align}
Now, defining
\begin{align}\label{eq:f_z}
	z(i)=\left\{\begin{array}{lr}
		\mu(u(i)-p(i)),&\text{if }u(i)\geq p(i),\\
		\frac{1}{\beta}(u(i)-p(i)),&\text{otherwise,}
	\end{array}\right.
\end{align}
we can re-write (\ref{eq:actual_energy_level}) as
\begin{align}\label{eq:actual_energy_level_w}
	E(i)=\min\Big\{\big[E(i-1)+z(i)\big]^{+},E_{\max}\Big\}.
\end{align}
We can further characterize the available space in the battery as
\begingroup
\allowdisplaybreaks
\begin{align}
	\thickbar{E}(i)&=E_{\max}-E(i)\nonumber\\
	&=E_{\max}-\min\Big\{\big[E(i-1)+z(i)\big]^{+},E_{\max}\Big\}\nonumber\\
	&=\Big[E_{\max}-\big[E(i-1)+z(i)\big]^{+}\Big]^{+}\nonumber\\
	&=\Big[\min\big\{E_{\max}-E(i-1)-z(i),E_{\max}\big\}\Big]^{+}\nonumber\\
	&=\Big[\min\big\{\thickbar{E}(i-1)-z(i),E_{\max}\big\}\Big]^{+}\nonumber\\
	&=\min\Big\{\big[\thickbar{E}(i-1)-z(i)\big]^{+},E_{\max}\Big\}.\label{eq:actual_available_space}
\end{align}
\endgroup
We can easily observe the analogy between the actual energy level in (\ref{eq:actual_energy_level_w}) and the available space to store more energy in (\ref{eq:actual_available_space}). When $E(i-1)+z(i)\geq E_{\max}$ in (\ref{eq:actual_energy_level_w}), there is an energy overflow, and the system dissipates the excess energy. On the other hand, when $\thickbar{E}(i-1)-z(i)\geq E_{\max}$ in (\ref{eq:actual_available_space}), an energy outage happens, and the data transmission is not performed. The control of energy overflows is important in order to utilize the harvested energy as efficiently as possible. Therefore, one may consider to implement an energy demand process that consumes the harvested energy as quickly as possible, which in turn may lead to energy outages. However, the control of energy outages is also necessary in order to minimize the number of transmission interruptions. Furthermore, we also know that the battery life is extended if only a fraction of the battery capacity is used for charging and discharging, although the entire battery capacity is used for energy storage \cite{ghiassi2015toward}. Therefore, we can consider the feasible battery capacity, $E_{\text{c}}$, as the fraction of the storage capacity that is used for charging and discharging. Then, we can have the case, in which the energy level falls below $E_{\min}$ and the actual available space in the battery becomes more than $E_{\text{c}}$, i.e., $\thickbar{E}(i)>E_{\text{c}}$, as an energy underflow, and set the energy underflow probability, $\Pr\{\thickbar{E}(i)\geq E_{\text{c}}\}$, as the battery constraint. Hence, we can regard the energy demand process, $p(i)$, which avoids energy underflows as much as possible so that the battery life is extended and the number of transmission interruptions is minimized, as an investigation problem.

\subsection{Energy Underflow}\label{sec:energy_underflow}
In this section, we take advantage of the similarity between a single server queueing system and the aforementioned energy storage model in (\ref{eq:actual_available_space}), and invoke the large deviation principle and queueing theory \cite{Chang}. We know that given a single service provider in a queueing system with a first in-first out policy, the steady-state queue length tail distribution has a characteristic decay rate, i.e., the decay rate of the tail probability of the queue \cite{chang1994effective}. Using this decay rate, we obtain a simple exponential expression for the queue overflow probability (i.e., the probability that the queue is greater than a threshold), which is a function of the desired threshold and the characteristic decay rate, given that the queueing capacity is infinite. In practical systems, this exponential function approximates the queue overflow probability very closely when the queueing capacity and the desired threshold are very large. Similarly, we assume that the battery size is infinite in the model described in Section \ref{sec:system_model}, i.e., $E_{\max}=\infty$, and define the decay rate of the tail distribution of the available space in the battery, given in (\ref{eq:actual_available_space}), as follows.

\begin{definition}\label{def:definition_decay_rate}
Given stationary and ergodic energy arrival and demand processes, $u(i)$ and $p(i)$, the available space decay rate of the battery in the steady-state is defined as
\begin{equation}\label{eq:decay_rate}
	\theta\triangleq-\lim_{E_{\text{th}}\to\infty}\frac{\log\Pr\{\thickbar{E}\geq E_{\text{th}}\}}{E_{\text{th}}},
\end{equation}
under the stability condition\footnote{Because we want to minimize the number of transmission interruptions and extend the battery life, we need to stabilize $\thickbar{E}(i)$. Therefore, we should have $\mathbb{E}\{p(i)\}<\mathbb{E}\{u(i)\}$ or $0<\mathbb{E}\{z(i)\}$ as the stability condition.}, i.e., $0<\mathbb{E}\{z(i)\}$, where $E_{\text{th}}$ is the threshold for the available space in the battery, and the random variable, $\thickbar{E}$, corresponds to the steady-state distribution of the available space in the battery.
\end{definition}

We can easily see that the expression (\ref{eq:decay_rate}) suggests an approximation for the probability that the available space is greater than a given threshold with an exponential function of $\theta$ and $E_{\text{th}}$, given that the threshold and the battery capacity are very large. Specifically, we can approximate the energy underflow probability with an exponential function of the decay rate, $\theta$, and the feasible battery capacity, $E_{\text{c}}$, as $\Pr\{\thickbar{E}(i)\geq E_{\text{c}}\}\approx e^{-\theta E_{\text{c}}}$. We can also infer that smaller $\theta$ refers to an energy demand process that consumes the energy in the battery very quickly, and hence a high energy underflow probability, whereas larger $\theta$ refers to an energy demand process that utilizes the available energy in a moderate way, and hence a low energy underflow probability.

Now, noting that $E_{\max}=\infty$, we re-write the available space given in (\ref{eq:actual_available_space}) as
\begin{align*}
	\thickbar{E}(i)=\big[\thickbar{E}(i-1)-z(i)\big]^{+}.
\end{align*}
Then, by considering a work-conserving\footnote{We consider that the transmitter has always data to transmit. Therefore, it consumes a certain amount of energy as long as there is energy in the battery, and we regard the energy demand process as work-conserving. We have this assumption because the harvested energy is utilized for data transmission only, and the control of energy underflows becomes important for an efficient use of the harvested energy when there is data in the transmitter buffer. Otherwise, when there is no data in the buffer, the transmitter harvests energy until the battery becomes full, and then stops harvesting energy.} energy demand process, we have a unique $\theta^{\star}$ such that \cite[Theorem 2.1]{chang1995effective}
\begin{equation}\label{eq:balance_equation}
\lim_{t\to\infty}\frac{1}{t}\log_{\text{e}}\mathbb{E}_{z}\left\{e^{-\theta^{\star}\sum_{i=1}^{t}z(i)}\right\}=0,
\end{equation}
where $\theta^{\star}$ is the available space decay rate provided in (\ref{eq:decay_rate}). The expression (\ref{eq:balance_equation}) states that given energy arrival and demand processes, and an available space decay rate for a feasible battery capacity, we can adjust\footnote{We assume that we can control the energy demand process but not the energy arrival process.} the parameters of the energy demand process under the stability condition. Specifically, let us consider that we have a transmission system that operates under the energy underflow probability constraint, $\Pr\{\thickbar{E}(i)\geq E_{\text{c}}\}$:
\begin{enumerate}
	\item We characterize the available space decay rate using the relation $\theta\approx-\frac{\log\Pr\{\thickbar{E}(i)\geq E_{\text{c}}\}}{E_{\text{c}}}$.
	\item We plug $\theta$ into (\ref{eq:balance_equation}), and find the energy demand parameters that guarantees the defined equality.
\end{enumerate}
For instance, we can adjust the average energy demand rate or the peak energy demand rate. Basically, considering the battery characteristics, $\theta$ is a design parameter that tells about the transmission parameters.

\begin{remark}
The aforementioned exponential approximation holds when the feasible capacity is large. As for the cases, when the feasible capacity is small, we refer to \cite{elwalid1995fundamental,wu_negi}, and we can show that the energy underflow probability is expressed as
\begin{equation*}
	\Pr\{\thickbar{E}(i)\geq E_{\text{c}}\}\approx\delta e^{-\theta E_{\text{c}}},
\end{equation*}
where $\delta$ is the probability that the battery is full, i.e., $\delta=\Pr\{\thickbar{E}(i)=0\}$.
\end{remark}

\section{Performance Analysis}\label{sec:performance_analysis}
In this section, we substantiate our analytical findings with simulation results. In the following, we first provide the energy underflow probability results for two different energy demand policies. Subsequently, we describe the channel model we consider for data transmission, and compare the average data service rate results of different energy demand policies.

\subsection{Energy Underflow Probability}
\begin{figure}
	\centering
	\includegraphics[width=0.45\textwidth]{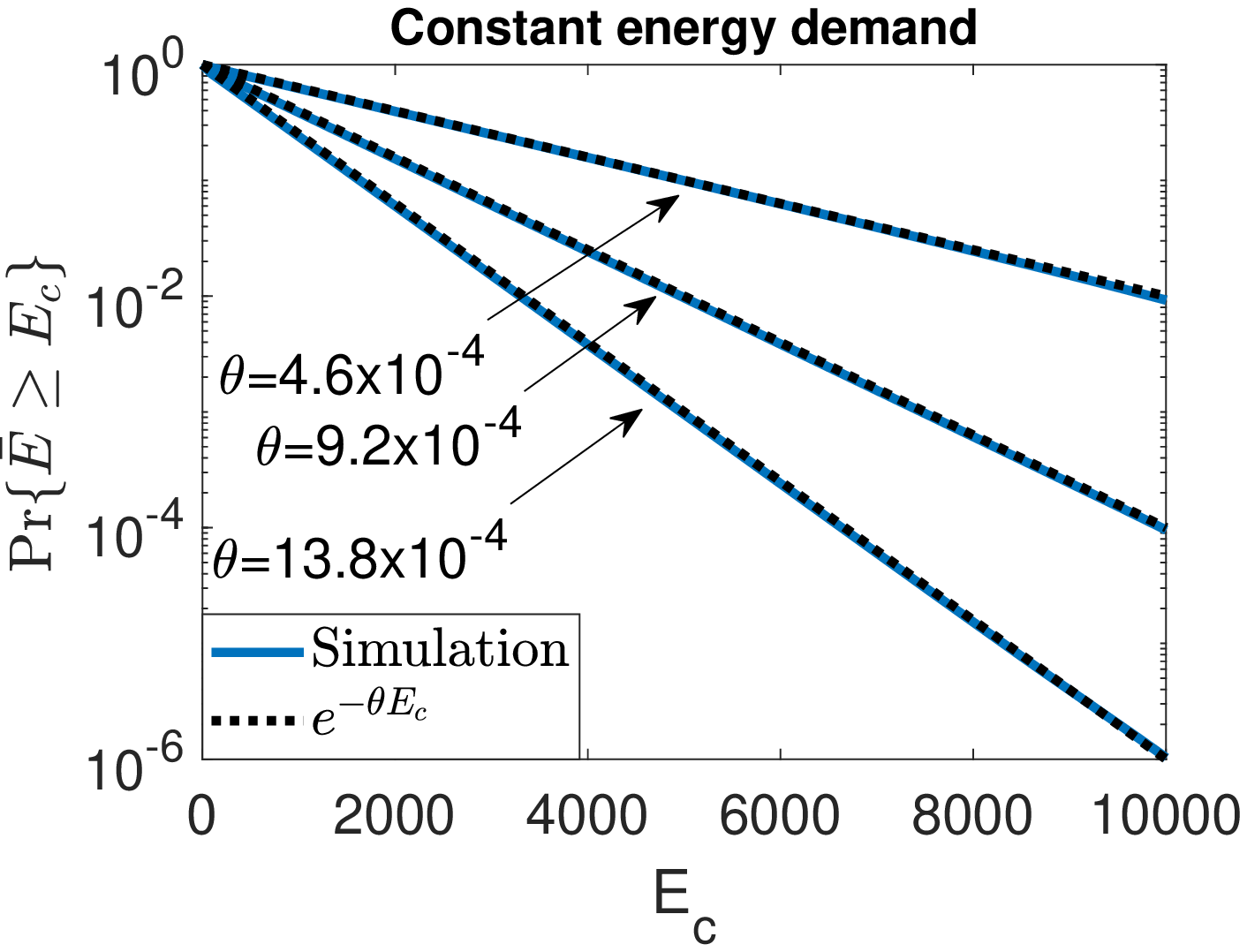}
	\caption{The energy underflow probability vs. the feasible battery capacity.}\label{fig:fig_1}
\end{figure}
Given that a constant-level energy demand process (a constant power allocation policy) exists, i.e., $p(i)=p$, we can express the probability density function of $z$ in (\ref{eq:f_z}) as
\begin{align*}
f_{z}(z)=\left\{\begin{array}{lr}
		\beta f_{u}(\beta z+p),&\text{if }-\frac{p}{\beta}\leq z\leq0,\\
		\frac{1}{\mu}f_{u}\left(\frac{z}{\mu}+p\right),&0\leq z,\\
		0,&\text{otherwise,}
	\end{array}\right.
\end{align*}
where $f_{u}(u)$ is the probability density function of the energy arrivals. Now, considering an energy harvesting process with exponentially distributed arrivals, i.e., $f_{u}(u)=\lambda_{u}e^{-\lambda_{u}u}$, we can express (\ref{eq:balance_equation}) for given $\theta>0$ as follows:
\begin{align}\label{eq:balance_equation_2}
\frac{\lambda_{u}\beta}{\lambda_{u}\beta+\theta}\left(e^{\frac{\theta p}{\beta}}-e^{-\lambda_{u}p}\right)+\frac{\lambda_{u}}{\lambda_{u}+\theta\mu}e^{-\lambda_{u}p}=1.
\end{align}
Above, $1/\lambda_{u}$ is the average energy arrival, and we assume that the energy arrivals are independent and identically distributed. Now, let $p^{\star}$ be the value that satisfies (\ref{eq:balance_equation_2}). As long as the constant energy demand value is less than $p^{\star}$, the desired energy underflow probability for a given feasible battery capacity, i.e., $\Pr\{\thickbar{E}(i)\geq E_{\text{c}}\}\approx e^{-\theta E_{\text{c}}}$, is sustained in the steady-state.

We plot the energy underflow probability results of the finite-size battery simulations in Fig. \ref{fig:fig_1}, and compare them with the exponential approximations. We set the battery capacity to 15000 energy units and the average energy arrival to 100 energy units, i.e., $E_{\max}=1.5\times10^{4}$ and $\lambda_{u}=0.01$, and consider the following available space decay rates: $\theta=4.6\times10^{-4}$, $9.2\times10^{-4}$ and $13.8\times10^{-4}$, which, for instance, correspond to $\Pr\{\thickbar{E}(i)\geq E_{\text{c}}\}\approx10^{-2}$, $10^{-4}$ and $10^{-6}$, respectively, for $E_{\text{c}}=10^{4}$. We set the feasible battery capacity range from $10^{3}$ to $10^{4}$ energy units given that $E_{\max}=1.5\times10^{4}$, because the expected number of equivalent full battery cycles is increased before the initial battery capacity falls below $80\%$ of the capacity \cite{guena2006depth}. We further set the battery charging and discharging rates to $0.85$ and $0.80$, i.e., $\mu=0.85$ and $\beta=0.80$, respectively. For more information on charging and discharging rates of different battery technologies, we refer interested readers to \cite{valoen2007effect}. As seen in Fig. \ref{fig:fig_1}, we can easily determine the energy underflow probability very closely with the exponential function approximation, and the energy underflow probability decreases with the increasing available space decay rate, $\theta$.

\begin{figure}
	\centering
	\includegraphics[width=0.45\textwidth]{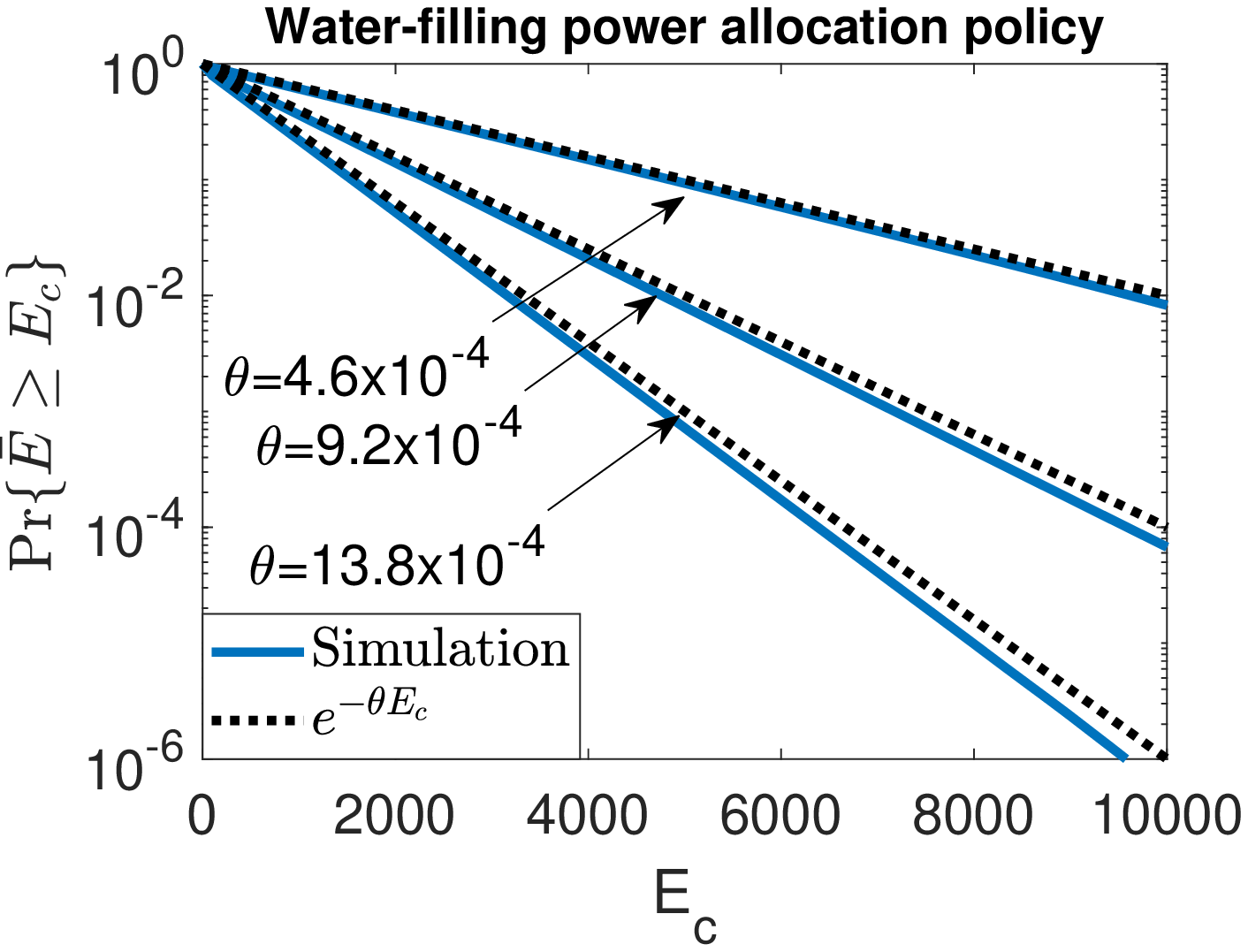}
	\caption{The energy underflow probability vs. the feasible battery capacity.}\label{fig:fig_2}
\end{figure}
We further employ the water-filling power allocation policy and set the energy demand to
\begin{align}
p(i)=N\sigma_{w}^{2}\left[\frac{1}{\varepsilon}-\frac{1}{h_{\text{pow}}(i)}\right]^{+},
\end{align}
where $h_{\text{pow}}(i)$ is the channel gain power in the $i^{\text{th}}$ time frame, and $\varepsilon$ is a cutoff value such that $\mathbb{E}_{z}\left\{e^{-\theta z(i)}\right\}=1$ for given $\theta$ and the given energy arrival process. We consider the same setting that we have in Fig. \ref{fig:fig_1}, and plot the energy underflow probability as a function of the feasible battery capacity in Fig. \ref{fig:fig_2}. We observe again that we can approximate the energy underflow probability with an exponential function very closely for the defined feasible capacity range. We also note that in both Fig. \ref{fig:fig_1} and Fig. \ref{fig:fig_2}, the energy underflow probability decreases with increasing $\theta$, because the battery constraints become more stringent, and the energy demand rates decrease in order to keep the energy level above $E_{\min}$ with increased probability.

\subsection{Average Data Service Rate}
\begin{figure}
	\centering
	\includegraphics[width=0.45\textwidth]{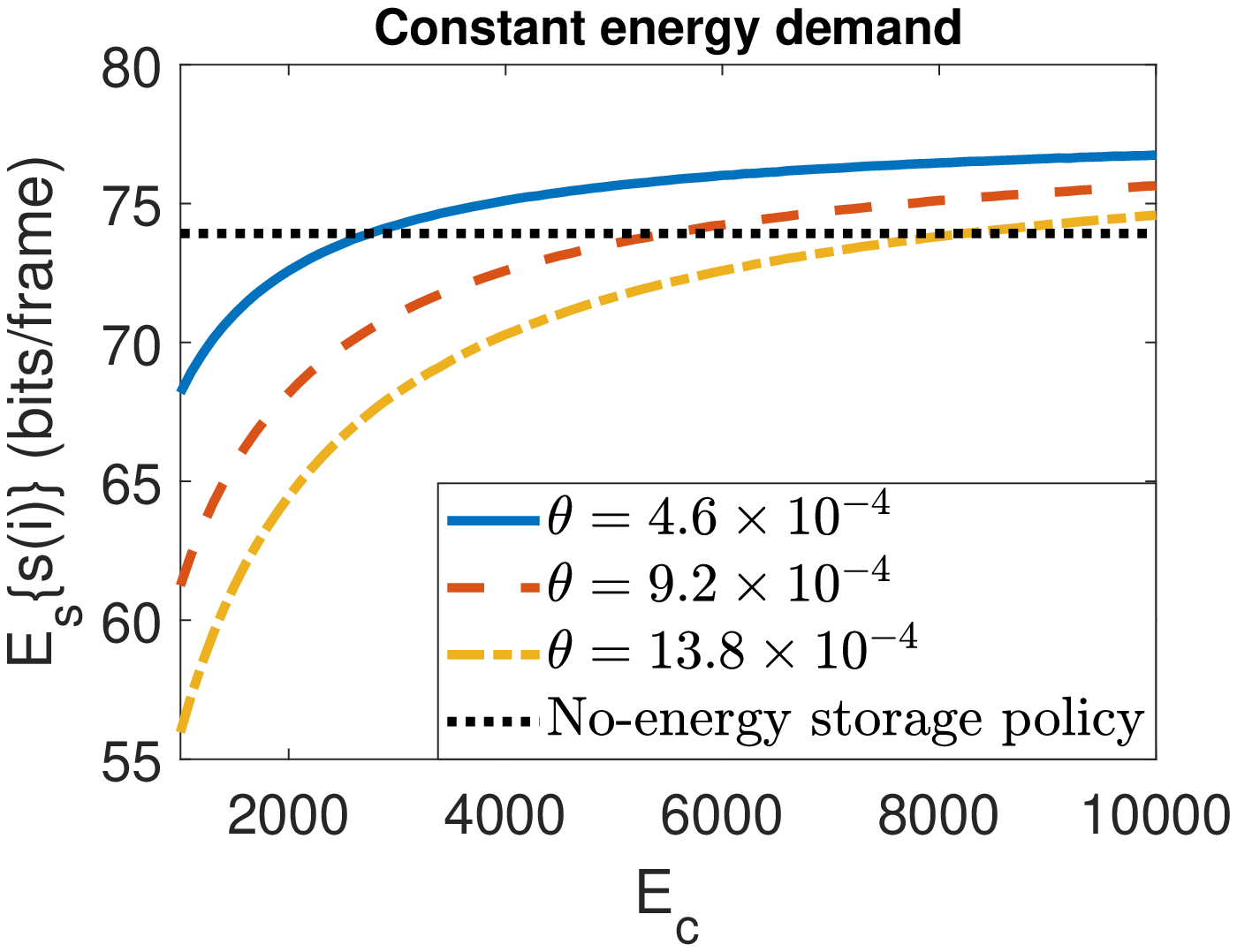}
	\caption{The average service rate vs. the feasible battery capacity.}\label{fig:fig_3}
\end{figure}
In order to investigate whether energy storage is useful when there are energy losses during charging and discharging, we compare the average data service rates in the wireless channel obtained when the energy storage is employed and when there is no energy storage. We assume that the transmitter receives $s(i)$ bits from the data source in each time frame. After encoding and modulating the received bits into a vector of $N$ symbols, i.e., $\mathbf{x}(i)$, the transmitter sends $\mathbf{x}(i)$ to the corresponding receiver through the wireless channel. During the transmission, the input-output relation in the channel is given as
\begin{equation}\label{eq:input_output}
\mathbf{y}(i)=\mathbf{x}(i)h(i)+\mathbf{w}(i),
\end{equation}
where $\mathbf{y}(i)$ and $\mathbf{x}(i)$ are the channel output and input vectors, respectively, and $\mathbf{w}(i)$ is the additive complex Gaussian noise vector with independent and identically distributed zero-mean samples. The covariance matrix of $\mathbf{w}(i)$ is $\mathbb{E}_{\mathbf{w}}\left\{\mathbf{w}(i)\mathbf{w}^{*}(i)\right\}=\sigma_{w}^{2}\mathbf{I}_{N\times N}$, where $\sigma_{w}^{2}$ is the average noise power, $\mathbf{w}^{*}(i)$ is the conjugate transpose of the noise vector, and $\mathbf{I}_{N\times N}$ is the $N\times N$ identity matrix. The channel input is constrained as follows: $||\mathbf{x}(i)||^{2}\leq p_{\text{c}}(i)$, where $p_{\text{c}}(i)$ is the amount of consumed energy in the $i^{\text{th}}$ time frame. When there is no energy storage, the consumed energy is equal to the instantaneous energy arrival, i.e., $p_{\text{c}}(i)=u(i)$. When the aforementioned energy storage is employed, the consumed energy becomes
\begin{align*}
p_{\text{c}}(i)=\left\{\begin{array}{lr}
		p(i),&\text{if }E(i-1)\geq\frac{[p(i)-u(i)]^{+}}{\beta},\\
		u(i)+\beta E(i-1),&\text{otherwise.}
	\end{array}\right.
\end{align*}
Moreover, given that a block-fading channel exists, the channel fading coefficient, $h(i)$, remains constant during one time frame, and changes independently from one frame to another. Furthermore, $N$ is relatively large; hence, the decoding error probability is negligible as long as the transmission rate is lower than or equal to instantaneous mutual information between the channel input $\mathbf{x}(i)$ and the channel output $\mathbf{y}(i)$. Therefore, given that the $h(i)$ is known by the transmitter, we set the transmission rate to the instantaneous mutual information. Principally, the number of  bits taken from the source, $s(i)$, is equal to the instantaneous mutual information in the $i^{\text{th}}$ time frame, i.e., $s(i)=\mathcal{I}(\mathbf{x}(i);\mathbf{y}(i))$. Hence, $s(i)$ is the data service rate in the channel.

\begin{figure}
	\centering
	\includegraphics[width=0.45\textwidth]{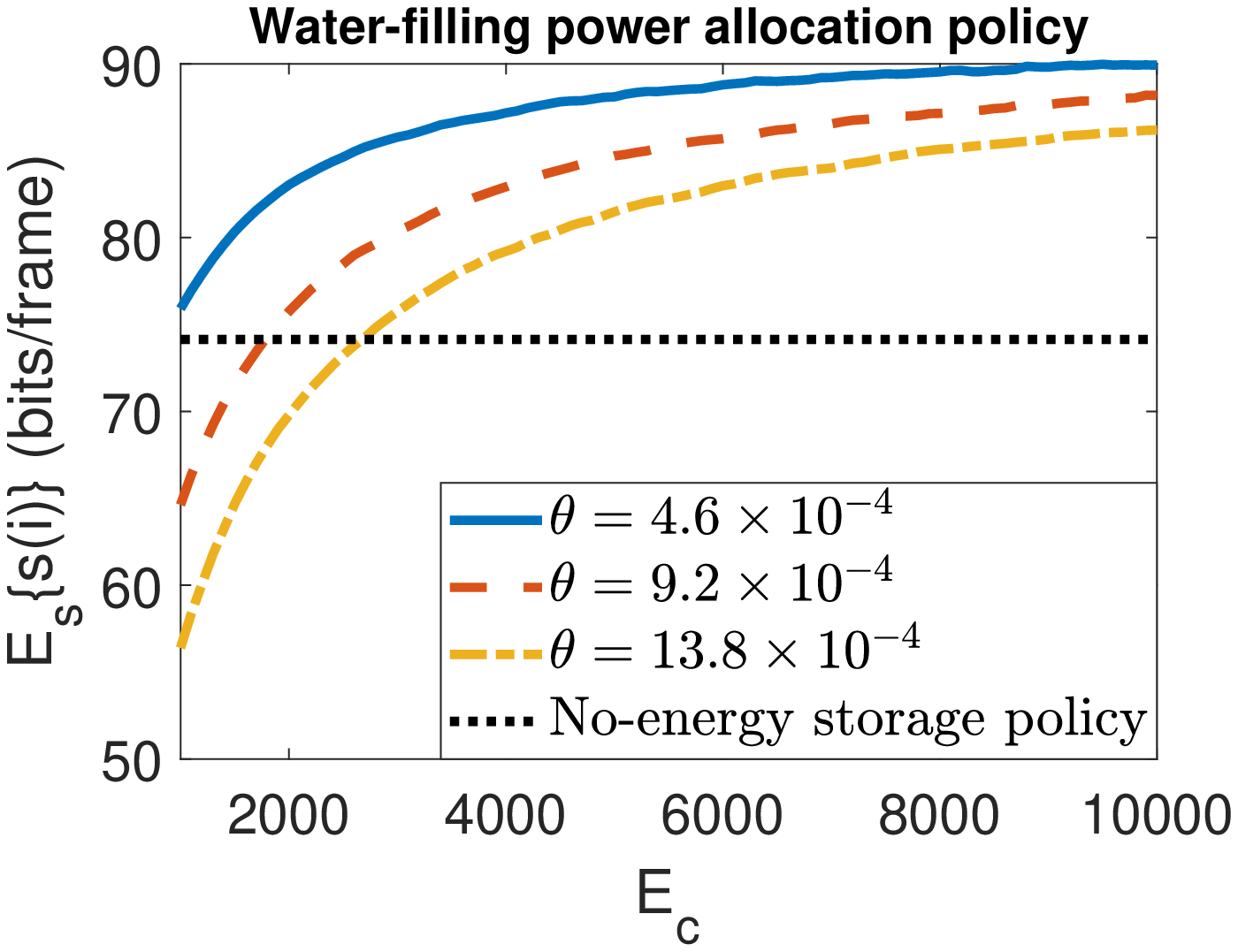}
	\caption{The average service rate vs. the feasible battery capacity.}\label{fig:fig_4}
\end{figure}
Given that the transmitter employs zero-mean, independent and identically complex Gaussian distributed channel inputs, we can express the average data service rate as\[\mathbb{E}_{s}\{s(i)\}=\mathbb{E}_{h_{\text{pow}}}\left\{N\log_{2}\left(1+\frac{p_{\text{c}}(i)h_{\text{pow}}(i)}{N\sigma_{w}^{2}}\right)\right\},\] where $h_{\text{pow}}(i)=|h(i)|^{2}$. We further consider that a sufficient averaging of the additive noise in one time frame is applicable, and perform a power allocation policy among the samples of $\mathbf{x}(i)=[x^{*}_{1}(i)\cdots x^{*}_{N}(i)]^{*}$ such that $\mathbb{E}\{||\mathbf{x}(i)||^{2}\}=p_{\text{c}}(i)$ and $x_{j}(i)\sim\mathcal{CN}(0,\frac{p_{\text{c}}(i)}{N})$. In Fig. \ref{fig:fig_3} and Fig. \ref{fig:fig_4}, we plot the average data service rates when the constant energy demand process is employed and when the water-filling power allocation based energy demand process is employed, respectively, for different energy underflow probability constraints, i.e., $\theta=4.6\times10^{-4}$, $9.2\times10^{-4}$ and $13.8\times10^{-4}$. The dotted black lines indicate the average data service rates obtained when there is no energy storage, which we refer to as \emph{no-energy storage policy}. When the feasible battery capacity is large, the average data service rates of the transmitter that employs energy storage are greater than the rates obtained by the transmitter that has no energy storage. However, with decreasing feasible battery capacity, the average service rates decrease and even fall below the rates obtained without energy storage. We further observe that the average data service rates increase with looser energy underflow probability constraints. Particularly, the average data service rates in both policies are higher with lower $\theta$ values. We finally note that the water-filling power allocation policy based energy demand process results in higher rates than the constant-level energy demand process does. More importantly, although there are energy losses during charging and discharging the battery, the energy storage policy is better than the no-energy storage policy when the feasible battery capacity is large enough.

\section{Conclusion}
In this paper, we have proposed an analytical framework for energy management in a wireless transmission system where the system harvests energy and stores it in a battery with energy losses during charging and discharging. We have initially elaborated energy underflows at the battery, and then obtained the energy underflow probability taking advantage of the large deviation principle. We have further characterized the available space decay rate parameter, and shown that we can use this parameter as a metric to indicate the relation between the energy arrival and demand processes. Using this relation, we have also described how we can adjust the energy demand parameters to sustain the energy underflow probability constraint at the battery. We have finally substantiated our analytical findings with numerical illustrations, and compared the average service rates of a transmitter with a battery that has energy losses during charging and discharging, and a transmitter that utilizes the energy as soon as it is harvested.

\bibliographystyle{IEEEtran}
\bibliography{IEEEabrv,References}

\end{document}